\documentclass[letterpaper,aps,prb,twocolumn,superscriptaddress,showpacs]{revtex4-1}

\usepackage{amsmath}
\usepackage{amssymb}
\usepackage{bm}

\usepackage{graphicx}
\usepackage[colorlinks=true,citecolor=blue,linkcolor=blue,urlcolor=blue]{hyperref}

\usepackage[T1]{fontenc}
\usepackage{textcomp}
\usepackage{txfonts}

\begin{document}

\title{Electroluminescence spectra in weakly coupled single-molecule junctions}

\author{J.~S.~Seldenthuis}
\author{H.~S.~J.~van~der~Zant}
\affiliation{Kavli Institute of Nanoscience, Delft University of Technology, Lorentzweg 1, 2628 CJ Delft, The Netherlands}
\author{M.~A.~Ratner}
\affiliation{Department of Chemistry, Northwestern University, 2145 Sheridan Road, Evanston, Illinois 60208-3113, USA}
\author{J.~M.~Thijssen}
\affiliation{Kavli Institute of Nanoscience, Delft University of Technology, Lorentzweg 1, 2628 CJ Delft, The Netherlands}

\date{\today}

\begin{abstract}
We have combined \emph{ab initio} quantum chemistry calculations with a rate-equation formalism to analyze electroluminescence spectra in single-molecule junctions, measured recently by several groups in scanning tunneling microscope setups. In our method, the entire vibrational spectrum is taken into account. Our method leads to good quantitative agreement with both the spectroscopic features of the measurements and their current and voltage dependence. Moreover, our method is able to explain several previously unexplained features. We show that in general, the quantum yield is expected to be suppressed at high bias, as is observed in one of the measurements. Additionally, we comment on the influence of the vibrational relaxation times on several features of the spectrum.
\end{abstract}


\maketitle

\section{Introduction}

Scanning tunneling microscopy (STM) has proven to be an invaluable tool for studying single molecules and their potential applications in nanoscale devices. In addition to measuring conductance properties, the electroluminescence spectrum, induced by the tunneling current can be studied.\cite{Coombs1988} Electroluminescence has been experimentally observed on clean metal surfaces,\cite{Berndt1991,Berndt1993,Uehara1999} films,\cite{Hoffmann2001,Hoffmann2002} nanoparticles,\cite{Nilius2000} and single molecules.\cite{Qiu2003,Dong2004,Cavar2005,Wu2008} Additionally, electroluminescence has been observed in lithographically defined three-terminal CdSe single-nanocrystal junctions.\cite{Gudiksen2005} In the case of single molecules, the simultaneous measurement of the electrical and optical behavior has the potential to greatly enhance our understanding of nanoscale junctions, and, through detailed analysis of the vibrational spectrum, to provide valuable insight into the conformational structure of single molecules in a junction.\cite{Ward2008}

\begin{figure}
    \begin{tabular}{cc}
        \includegraphics[width=1.6in]{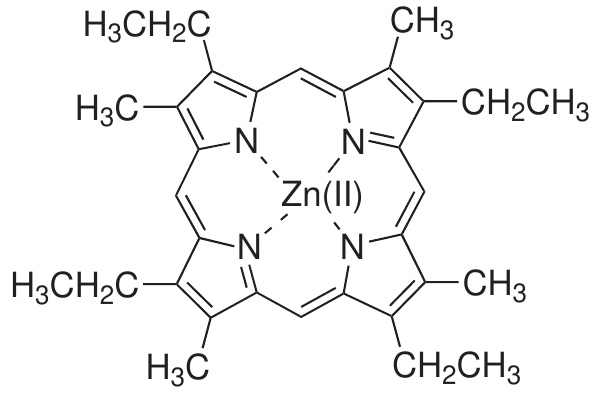} &
        \includegraphics[width=1.6in]{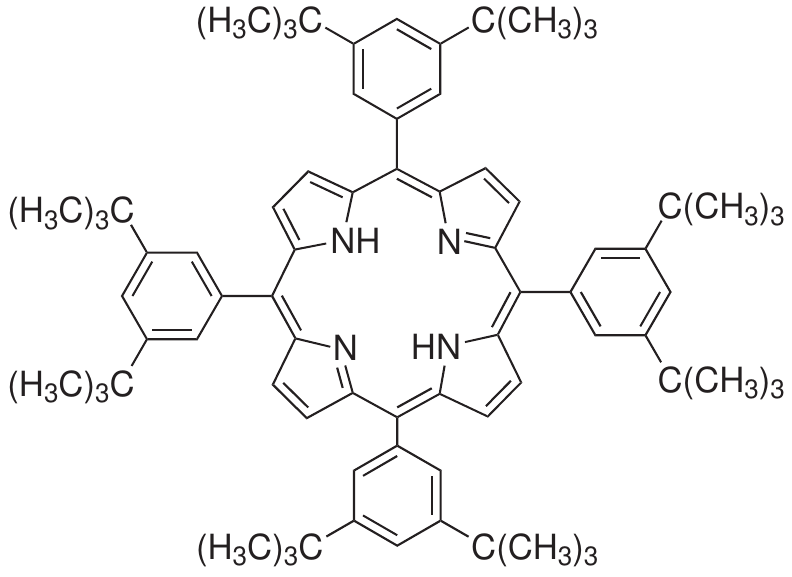} \\
        (a) & (b)
    \end{tabular}
    \caption{\label{fig:structure} Molecular structure of the isolated porphyrin derivatives used in the calculations: (a) ZnEtioI, measured by Qiu \emph{et al.}\ (Ref.~\onlinecite{Qiu2003}) and (b) H$_2$TBPP, measured by Dong \emph{et al.}\ (Ref.~\onlinecite{Dong2004}).}
\end{figure}

We have developed a method based on a combination of \emph{ab initio} density functional theory (DFT) calculations with a rate-equation formalism to analyze electroluminescence measurements on weakly coupled single-molecule junctions. In this paper we apply our method to two STM measurements on single porphyrin derivatives: Zn(II)-etioporphyrin~I [ZnEtioI, see Fig.~\ref{fig:structure}(a)] measured by Qiu \emph{et al.},\cite{Qiu2003} and \emph{meso}-tetrakis(3,5-diterbutylphenyl)porphyrin [H$_2$TBPP, see Fig.~\ref{fig:structure}(b)] measured by Dong \emph{et al.}\cite{Dong2004} It is known that the relative electronic coupling strengths to the source (S) (STM tip) and drain (D) (substrate) electrodes greatly affect molecular electroluminescence,\cite{Buker2002} and in order to suppress quenching due to the substrate, it is necessary to employ a spacer layer. Qiu \emph{et al.}\cite{Qiu2003} have used a thin Al$_2$O$_3$ film on the NiAl(110) substrate to act as a spacer layer. Although the electronic coupling to the leads is still highly asymmetric, electroluminescence is sufficiently enhanced to be observable. Dong \emph{et al.},\cite{Dong2004} on the other hand, have deposited several monolayers of H$_2$TBPP, resulting in a nearly symmetric coupling of the molecule to the leads. Since these measurements investigate similar molecules in different regimes (asymmetric and symmetric coupling), they provide a good test case to understand the physics of single-molecule electroluminescence.

It is known that plasmons in the substrate and the STM tip can mediate photoemission, and in the case of surface enhanced Raman scattering this can even be a powerful spectroscopic tool.\cite{Nie1997} However, in order to observe electroluminescence originating solely from a single molecule, the signal from plasmons should be suppressed as much as possible. In the experiments under discussion, several measures have been taken to this end, including using atomically flat substrates and tungsten STM tips. Also, in the measurements of Qiu \emph{et al.},\cite{Qiu2003} applying a series of high-voltage pulses between the tip and the substrate seems to make the plasmon spectrum smoother. Although an enhancement of the photoemission rate due to the STM tip and the substrate probably still exists, the experiments show that this enhancement is generally rather structureless and does not affect the shape of the spectra. Moreover, control measurements in both groups show that the electroluminescence intensity of the bare surface is much weaker than that of the molecules.\cite{Qiu2003,Dong2004} We will therefore ignore the effect of plasmons from now on.

The measurement of Qiu \emph{et al.}\cite{Qiu2003} has been previously analyzed by Buker and Kirczenow using the Lippmann-Schwinger Green's function scattering technique.\cite{Buker2005,Buker2008} In this paper we employ the rate-equation formalism, which allows us to take the entire vibrational spectrum of the molecule into account.

\section{Method}\label{sec:Method}

\begin{figure}
    \includegraphics{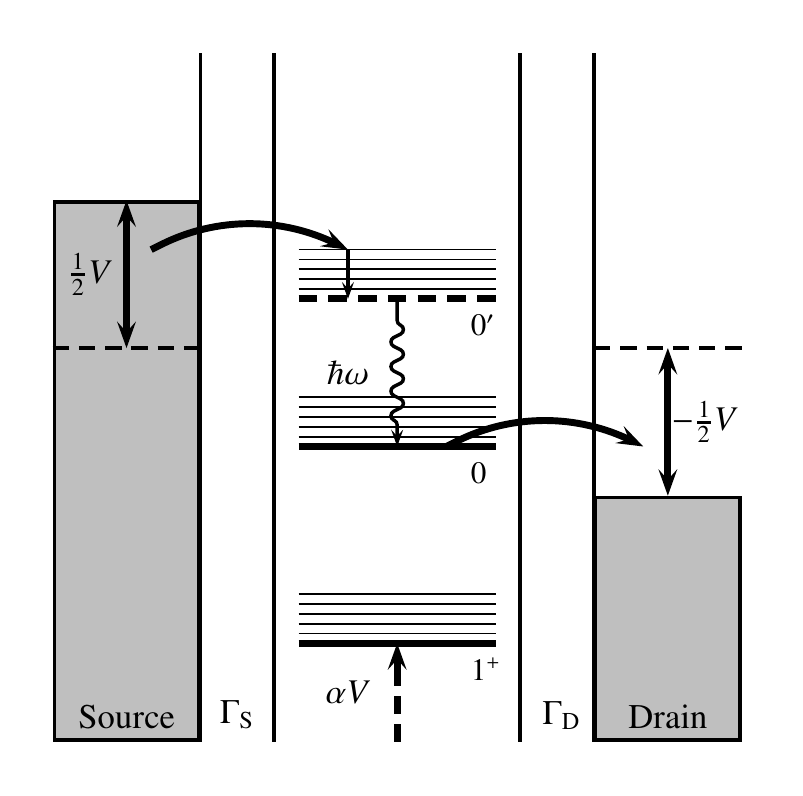}
    \caption{\label{fig:set} Schematic picture of sequential tunneling through a molecular junction. $\Gamma_\text{S}$ and $\Gamma_\text{D}$ are the electronic couplings to the source and drain electrodes, respectively, and $\alpha$ is the electrostatic coupling to the bias voltage. These quantities can, in principle, be different for different orbitals. The thick solid lines represent the HOMO of the $1^+$ and neutral charge state, and the dashed line represents the LUMO of the neutral charge state. The thin lines are vibrational excitations. The small vertical arrow at the top indicates vibrational relaxation, and the wavy line indicates the emission of a photon.}
\end{figure}

The magnitude of the electronic coupling constants obtained from the total current and the $dI$/$dV$-curves in the measurements (discussed in Sec.~\ref{sec:ZnEtioI}) is indicative of Coulomb blockade [see Fig.~2 in Ref.~\onlinecite{Qiu2003}, and Fig.~\ref{fig:ZnEtioI_didv}(a)], suggesting that the molecule-electrode coupling is weak ($\Gamma_\text{S},\Gamma_\text{D},k_{\text{B}}T\ll\Delta E$), and the electron addition energies ($\Delta E$) only allow the tunneling of one electron at a time (sequential tunneling). A schematic picture of this process is shown in Fig.~\ref{fig:set}.\cite{Note1} In this picture, current can flow as soon as the highest-occupied molecular orbital (HOMO) of the neutral charge state enters the bias window. When also the lowest-unoccupied molecular orbital (LUMO) becomes available, it is possible for the molecule to be in an electronically excited state when an electron tunnels onto the LUMO instead of the HOMO. If the coupling to the drain ($\Gamma_\text{D}$) is weak enough, and the electron stays on the molecule for some time, the excited state can decay to the ground state and emit a photon. The efficiency of this process is given by the luminescence quantum yield, which is defined as the number of emitted photons per transmitted electron.

\begin{figure}
    \includegraphics{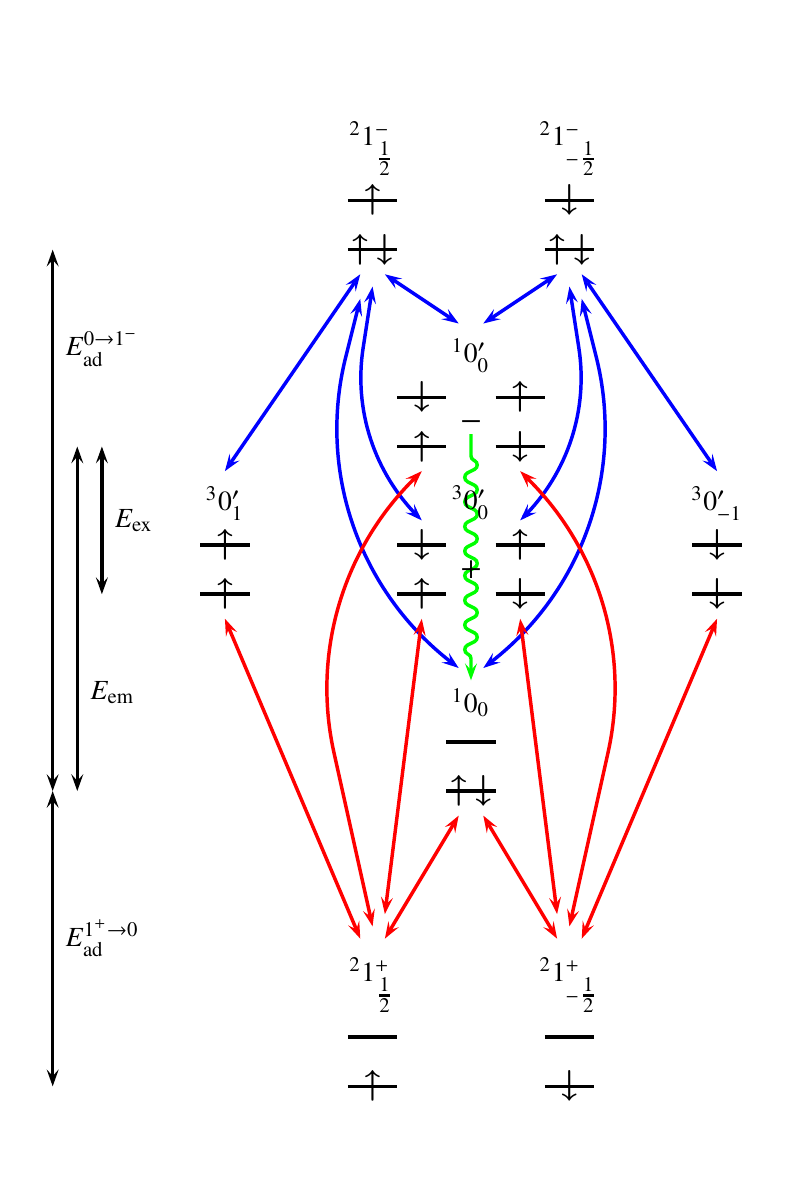}
    \caption{\label{fig:states} The one-, two-, and three-electron states of a single molecule, with an even number of electrons in the neutral state and a non-degenerate HOMO and LUMO. Vibrational excitations are not shown. The labels are of the format $^{2S+1}Q_{m_s}$, where $Q$ is the charge of the molecule, $S$ the total spin, and $m_s$ the eigenvalue of $S_z$. Electronically excited states are denoted by $'$. Transitions from the $1^+$ to neutral state are shown in red, transitions from the neutral to $1^-$ in blue, and the photoemission transition is shown in green. $E_\text{ad}^{Q\rightarrow Q-1}$ is the electron addition energy, $E_\text{em}$ the photo-emission energy, and $E_\text{ex}$ the exchange energy between the singlet and triplet state of $0'$. Note that the energy differences are not to scale.}
\end{figure}

In the weak-coupling regime, the current-voltage characteristics and the electroluminescence spectra can be calculated with the rate-equation formalism.\cite{Beenakker1991,Boese2001,Bonet2002,McCarthy2003,Braig2003,Mitra2004,Koch2004,Wegewijs2005,Chang2007,Seldenthuis2008} In this formalism, a finite number of molecular states is taken into account, and all processes are described in terms of transitions between these states at a certain rate. The central quantity in this formalism is the vector of occupation probabilities $P_n$. Here $n$ is taken to be a general quantum number describing charge, spin, and electronic and vibrational excitations. The time evolution of the occupation probabilities is governed by the master equation,
\begin{equation}
\frac{dP_n}{dt}=\sum_{n'\neq n}\left(P_{n'}W_{n'\rightarrow n}-P_nW_{n\rightarrow n'}\right),
\end{equation}
where $W_{n'\rightarrow n}$ and $W_{n\rightarrow n'}$ are the transition rate constants between the states $n$ and $n'$ (given by Fermi's golden rule). This equation can be written in matrix-vector form,
\begin{equation}
\frac{d{\bm P}}{dt}={\bm W}{\bm P},
\end{equation}
where ${\bm W}$ is the rate-constant matrix with elements
\begin{equation}
W_{ij}=
    \begin{cases}
        -\sum_{k\neq i}W_{i\rightarrow k} & \text{if $i=j$,} \\
        W_{j\rightarrow i} & \text{otherwise.}
    \end{cases}
\end{equation}
The stationary states, with $\frac{d{\bm P}}{dt}=0$, correspond to the null space of ${\bm W}$, with the condition that all elements of ${\bm P}$ are non-negative, and $\sum_n P_n=1$.

The states of a single molecule with an even number of electrons in the neutral state, and a non-degenerate HOMO and LUMO are shown in Fig.~\ref{fig:states}. Ignoring vibrational excitations for the moment, the spin multiplicities of three charge states and one excited state result in a total of nine states, and therefore a $9\times 9$ rate-constant matrix. However, assuming states with the same total spin have equal electronic ($\Gamma_\text{S}$ and $\Gamma_\text{D}$) and electrostatic ($\alpha$) (Ref.~\onlinecite{Note2}) couplings to the leads, and therefore equal transition rates, $^21_{\tfrac{1}{2}}^+$ and $^21_{-\tfrac{1}{2}}^+$ can be combined into $^21^+$, $^30_1'$, $^30_0'$, and $^30_{-1}'$ into $^30'$, and $^21_{\tfrac{1}{2}}^-$ and $^21_{-\tfrac{1}{2}}^-$ into $^21^-$, leaving only five states. Since we are now describing the system in terms of combined states, the transition rates differ from the original ones in that they have to be multiplied by a prefactor depending on the multiplicity of the states involved. Calling $M_n=2S_n+1$ the multiplicity of state $n$, the prefactor for the transition rate constant $W_{n\rightarrow n'}$ in this system is given by
\begin{equation}
C_{nn'}=
    \begin{cases}
        \frac{M_{n'}}{M_n} & \text{if $M_{n'}>M_n$,} \\
        1 & \text{otherwise}.
    \end{cases}
\end{equation}
This also carries over into the case where vibrational excitations are taken into account, as long as they are spin-independent.

The transition rates themselves depend on the type of transition: tunneling, photo-emission, or vibrational relaxation (indicated by the arrows in Fig.~\ref{fig:set}). Tunneling of electrons to and from the source and drain is responsible for transitions between states with a different number of electrons on the molecule. Here we only consider single-electron tunneling events. In the case of charging, where state $n'$ has one more electron than state $n$, the rate constant is given by $W_{n\rightarrow n'}=W_{n\rightarrow n'}^\text{S}+W_{n\rightarrow n'}^\text{D}$, where
\begin{equation}
W_{n\rightarrow n'}^\text{S}=C_{nn'}F_{nn'}\frac{\Gamma_\text{S}}{\hbar}f\left(E_\text{ad}^{n\rightarrow n'}-\left(\tfrac{1}{2}-\alpha\right)V\right),
\end{equation}
and correspondingly for $W_{n\rightarrow n'}^\text{D}$, where $F_{nn'}$ are the Franck-Condon factors (the overlap between the nuclear wave functions) (Refs.~\onlinecite{Seldenthuis2008} and \onlinecite{Ruhoff2000}) and $f$ is the Fermi function. Similarly, in the case of discharging, the transition rate constant is given by $W_{n'\rightarrow n}=W_{n'\rightarrow n}^\text{S}+W_{n'\rightarrow n}^\text{D}$, where
\begin{equation}
W_{n'\rightarrow n}^\text{S}=C_{n'n}F_{n'n}\frac{\Gamma_\text{S}}{\hbar}\left[1-f\left(E_\text{ad}^{n\rightarrow n'}-\left(\tfrac{1}{2}-\alpha\right)V\right)\right],
\end{equation}
and correspondingly for $W_{n'\rightarrow n}^\text{D}$. Photo-emission is possible from the singlet excited state $^10_0'$ to the ground state $^10_0$ (fluorescence). Transitions from the triplet states $^30_1'$, $^30_0'$, and $^30_{-1}'$ (phosphorescence) are spin forbidden, and can only occur in the presence of spin-orbit coupling. Since phosphorescence occurs on a much slower time scale than fluorescence --- and especially considering the fact that excited electrons can tunnel off the molecule in a junction --- we will only take photo-emission from the singlet state into account. The radiative transition rate constant from the excited state $n'$ to the ground state $n$ is given by\cite{Schatz1993}
\begin{equation}
W_{n'\rightarrow n}^\text{E}=F_{n'n}\frac{\omega^3}{3\pi\epsilon_0\hbar c^3}\left|\mu\right|^2,
\end{equation}
where $\omega$ is the frequency of the emitted light (determined approximately by the HOMO-LUMO gap), and $\mu$ is the transition dipole moment. Note that we are only considering radiative transitions between the excited state and the ground state. Since the time scale of both the radiative and non-radiative transitions (microsecond to nanosecond) is typically much slower than the charging and discharging time scale (approximately picosecond), these transitions will have a negligible effect on the occupation probabilities. Only the radiative transitions are directly observable, while the non-radiative transitions will usually have an imperceptible effect on the total current. We will therefore only include the former in the rate equations and ignore the latter. Vibrational relaxation is taken into account with a single relaxation time for all vibrationally excited states.\cite{Koch2004} Separating the vibrational quantum number $\nu$ from the other quantum numbers $n$, the transition rate constant for this relaxation is given by
\begin{equation}
W_{n\nu'\rightarrow n\nu}^\text{R}=\frac{1}{\tau}P_{n\nu}^\text{eq},
\end{equation}
where $\tau$ is the relaxation rate and
\begin{equation}
P_{n\nu}^\text{eq}=\frac{e^{-\frac{E_{n\nu}}{k_B T}}}{\sum_{\nu''}e^{-\frac{E_{n\nu''}}{k_B T}}}
\end{equation}
is the equilibrium occupation of the vibrational excitations according to the Boltzmann distribution. Note that a single relaxation time for all vibrational excitations is only a good approximation when relaxation is either much faster or much slower than all other rates. We will come back to this in Sec.~\ref{sec:H2TBPP}.

For the molecules under investigation, we have calculated the equilibrium geometry and the vibrational modes of the isolated system using DFT.\cite{Note3} Although DFT is well suited for ground-state calculations, taking excited states into account can be problematic. We have therefore approximated the excited state by forcing the molecule into a high-spin ($S=1$) configuration. Since the equilibrium geometry and the normal modes only depend on the occupation of the molecular orbitals through the charge density, and not on the total electron spin, this is not expected to have a significant effect on the vibrational spectrum. The Franck-Condon factors have been calculated from the equilibrium geometries and the normal modes by using the method of Ruhoff and Ratner,\cite{Ruhoff2000} while the transition dipole moments have been calculated with time-dependent DFT. This leaves the electronic and electrostatic couplings, and the position of the Fermi level within the HOMO-LUMO gap as fit-parameters. Note that these parameters depend on the unknown contact geometry and vary from sample to sample in the measurements. Since extracting the HOMO-LUMO gap from DFT calculations is delicate, this value is fitted to the measurements as well. These fit-parameters only have a small influence on the vibrational features of the spectrum, which is the focus of our research.

\begin{figure}
    \begin{tabular}{cc}
        \includegraphics{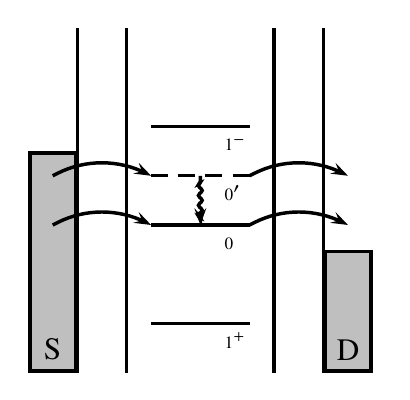} &
        \includegraphics{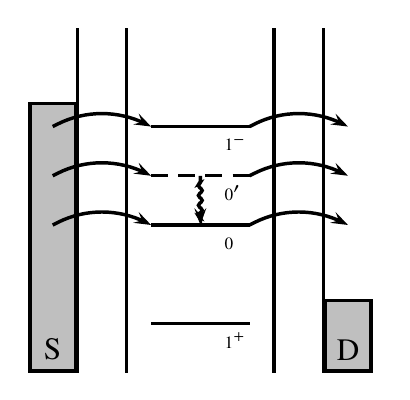} \\
        (a) & (b)
    \end{tabular}
    \caption{\label{fig:rates} (a) Sequential tunneling with one charge state ($0$) and one excited state ($0'$) in the bias window. (b) Sequential tunneling with two charge states ($0$ and $1^-$) and one excited state ($0'$) in the bias window. The horizontal lines represent transitions between many-electron states.}
\end{figure}

Before applying our method to the ZnEtioI and H$_2$TBPP porphyrin derivatives, let us first consider simple situations which can be solved analytically. We ignore vibrational excitations for the time being. In Fig.~\ref{fig:rates}(a), the $0$ and $0'$ states are in the bias window and available for transport, while the $1^-$ state remains unoccupied. In Fig.~\ref{fig:rates}(b), also the $1^-$ state is available. In both cases we assume the electronic and electrostatic couplings to the leads to be equal for all states, and the temperature to be low enough for the Fermi functions to be either 1 or 0. Since both ZnEtioI and H$_2$TBPP have a (nearly) degenerate LUMO, we double the multiplicities of the states involving the LUMO, resulting in the following five states: $^21^+$, $^10$, $^60'$, $^20'$, and $^41^-$. As a final approximation, we assume the photo-emission rate to be much smaller than the charging and discharging rates. It will therefore have a negligible effect on the occupation probability of the singlet excited state and can be omitted from the master equation. Since the rates for the singlet and triplet excited states are now equal, they can be combined into one excited state $^80'$, provided they are both in the bias window, leaving only four states to consider: $^21^+$, $^10$, $^80'$, and $^41^-$.

In the case of Fig.~\ref{fig:rates}(a), where the $^41^-$ state can be ignored, the rate-constant matrix is:
\begin{equation}
{\bm W}=\frac{1}{\hbar}
    \begin{pmatrix}
        -5\Gamma_\text{S} & 2\Gamma_\text{D} & \Gamma_\text{D} \\
        \Gamma_\text{S} & -2\Gamma_\text{D} & 0 \\
        4\Gamma_\text{S} & 0 & -\Gamma_\text{D}
    \end{pmatrix},
\end{equation}
resulting in the stationary occupation probability
\begin{equation}
{\bm P}=\frac{1}{9\Gamma_\text{S}+2\Gamma_\text{D}}
    \begin{pmatrix}
        2\Gamma_\text{D} \\
        \Gamma_\text{S} \\
        8\Gamma_\text{S}
    \end{pmatrix}.
\end{equation}
The current is given by
\begin{equation}
I=\frac{e}{\hbar}\Gamma_\text{S}5P_{1^+}=\frac{e}{\hbar}\frac{10\Gamma_\text{S}\Gamma_\text{D}}{9\Gamma_\text{S}+2\Gamma_\text{D}},
\end{equation}
and the electroluminescence intensity by
\begin{equation}
L=\tfrac{1}{4}\frac{\Gamma_\text{E}}{\hbar}P_{0'}=\frac{1}{\hbar}\frac{2\Gamma_\text{S}\Gamma_\text{E}}{9\Gamma_\text{S}+2\Gamma_\text{D}},
\end{equation}
where $\Gamma_\text{E}=\hbar W_\text{E}$, and the factor $\tfrac{1}{4}$ is due to the fact that only the singlet, \emph{i.e.}, one quarter of the states in $^80'$, can emit a photon. This equation shows that electroluminescence can indeed be quenched when the coupling to the drain is too large. The quantum yield is then
\begin{equation}\label{eq:q_a}
Q=e\frac{L}{I}=\tfrac{1}{5}\frac{\Gamma_\text{E}}{\Gamma_\text{D}}.
\end{equation}

In the case of Fig.~\ref{fig:rates}(b), the rate-constant matrix is:
\begin{equation}
{\bm W}=\frac{1}{\hbar}
    \begin{pmatrix}
        -5\Gamma_\text{S} & 2\Gamma_\text{D} & \Gamma_\text{D} & 0 \\
        \Gamma_\text{S} & -4\Gamma_\text{S}-2\Gamma_\text{D} & 0 & \Gamma_\text{D} \\
        4\Gamma_\text{S} & 0 & -\Gamma_\text{S}-\Gamma_\text{D} & 2\Gamma_\text{D} \\
        0 & 4\Gamma_\text{S} & \Gamma_\text{S} & -3\Gamma_\text{D}
    \end{pmatrix},
\end{equation}
resulting in the stationary occupation probability
\begin{equation}
{\bm P}=\frac{1}{\left(4\Gamma_\text{S}+\Gamma_\text{D}\right)\left(\Gamma_\text{S}+2\Gamma_\text{D}\right)}
    \begin{pmatrix}
        2\Gamma_\text{D}^2 \\
        \Gamma_\text{S}\Gamma_\text{D} \\
        8\Gamma_\text{S}\Gamma_\text{D} \\
        4\Gamma_\text{S}^2
    \end{pmatrix}.
\end{equation}
The current is now given by
\begin{align}
I & =\frac{e}{\hbar}\Gamma_\text{S}\left(5P_{1^+}+4P_0+P_{0'}\right) \nonumber \\
& =\frac{e}{\hbar}\frac{2\Gamma_\text{S}\Gamma_\text{D}\left(6\Gamma_\text{S}+5\Gamma_\text{D}\right)}{\left(4\Gamma_\text{S}+\Gamma_\text{D}\right)\left(\Gamma_\text{S}+2\Gamma_\text{D}\right)},
\end{align}
and the electroluminescence intensity by
\begin{equation}
L=\tfrac{1}{4}\frac{\Gamma_\text{E}}{\hbar}P_{0'}=\frac{2\Gamma_\text{S}\Gamma_\text{D}\Gamma_\text{E}}{\left(4\Gamma_\text{S}+\Gamma_\text{D}\right)\left(\Gamma_\text{S}+2\Gamma_\text{D}\right)},
\end{equation}
resulting in a quantum yield of
\begin{equation}\label{eq:q_b}
Q=\frac{\Gamma_\text{E}}{6\Gamma_\text{S}+5\Gamma_\text{D}}.
\end{equation}
Comparing Eqs.~\ref{eq:q_a} and \ref{eq:q_b} shows that the availability of a second charge state in the bias window changes the quantum yield by a factor of $\frac{\Gamma_\text{D}}{1.2\Gamma_\text{S}+\Gamma_\text{D}}$, \emph{i.e.}, it always decreases, with the magnitude of the change being determined by the asymmetry in the coupling. Although the expression for the change in the quantum yield depends on the particular multiplicities of the states involved, this is a general result, and can be easily understood. The availability of the second charge state provides a new non-radiative path for the excited state to decay (via the tunneling of an electron onto the partially occupied HOMO). This will decrease the probability of the molecule to be in the excited state, and therefore decrease the electroluminescence. At the same time, the extra charge state also provides what is effectively an extra conductance channel, thereby increasing the current. Both effects will decrease the quantum yield. We will come back to this when discussing the results for H$_2$TBPP in Sec.~\ref{sec:H2TBPP}.

\section{Results}

\subsection{ZnEtioI}\label{sec:ZnEtioI}

\begin{figure}
    \begin{tabular}{cc}
        \includegraphics{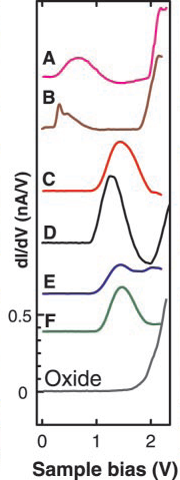} &
        \includegraphics{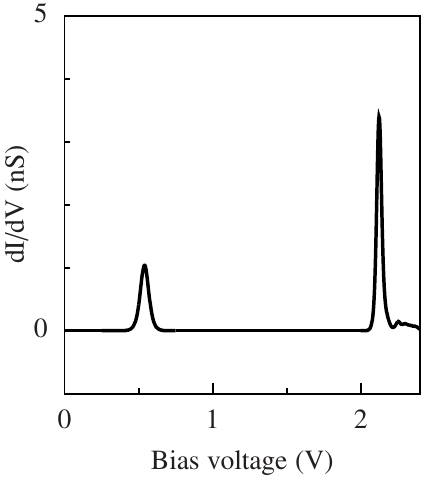} \\
        (a) & (b)
    \end{tabular}
    \caption{\label{fig:ZnEtioI_didv} (a) Measured conductance of six separate single ZnEtioI molecules (A--F) and the oxide surface. [From Qiu \emph{et al.}\ (Ref.~\onlinecite{Qiu2003}) reprinted with permission from AAAS.] (b) Calculated conductance of molecule A.  The electronic coupling is highly asymmetric ($\Gamma_\text{S}=0.637$~$\mu$eV and $\Gamma_\text{D}=10$~meV) and $T=77$~K.}
\end{figure}

The equilibrium configuration of the ZnEtioI molecule obtained from DFT calculations is shown in Fig.~\ref{fig:structure}(a), where the $\mathcal{S}_4$ ``saddle shape'' structure of the isolated molecule is in agreement with the STM topography analysis of Qiu \emph{et al.}\cite{Qiu2004} Figure~\ref{fig:ZnEtioI_didv} shows the measured and calculated conductance of the molecule in a STM junction. Qiu \emph{et al.}\ have measured the conductance of six separate single ZnEtioI molecules [A--F in Fig.~\ref{fig:ZnEtioI_didv}(a)], of which only A and B were observed to luminesce. These molecules have in common that their conductance plots show two peaks within the bias window. This is consistent with our rate-equation model, since both the HOMO and the LUMO have to be within the bias window for electroluminescence to be possible, and each will give rise to a peak in the conductance. The calculated conductance for molecule A is shown in Fig.~\ref{fig:ZnEtioI_didv}(b). Since only broadening due to temperature is taken into account in our rate-equation approach, and not due to the coupling to the leads, the calculated conductance peaks are much sharper than the measured peaks. It is striking that the relative peak heights of the HOMO and the LUMO are nearly identical to the measurement, even though the HOMO and LUMO are assumed to couple equally to the leads in the calculation. The difference in peak height is caused solely by the degeneracy of the LUMO and the multiplicity of the excited state.

\begin{figure}
    \begin{tabular}{cc}
        \multicolumn{2}{c}{\includegraphics{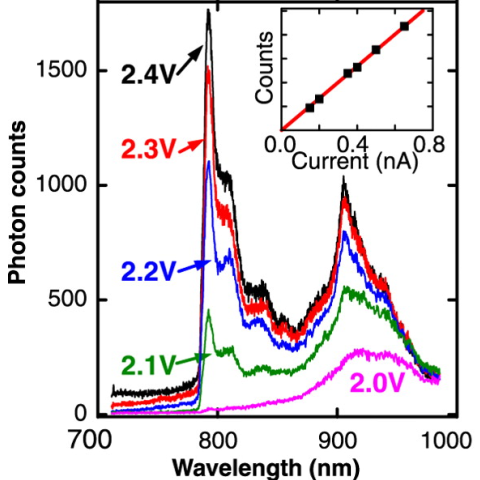}} \\
        \multicolumn{2}{c}{(a)} \\
        \includegraphics{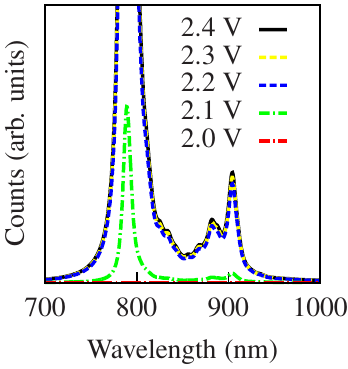} &
        \includegraphics{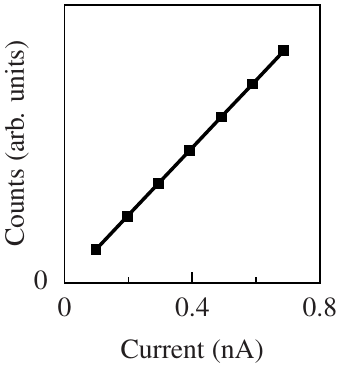} \\
        (b) & (c)
    \end{tabular}
    \caption{\label{fig:ZnEtioI_spectrum} (a) Measured electroluminescence spectrum of ZnEtioI as a function of bias voltage. [From Qiu \emph{et al.}\ (Ref.~\onlinecite{Qiu2003}) reprinted with permission from AAAS.] (b) Calculated spectrum as a function of bias voltage (77~K, assuming instantaneous vibrational relaxation). (c) Calculated current dependence of the 790~nm peak intensity at 2.35~V. Compare to the inset in (a).}
\end{figure}

\begin{figure}
    \begin{tabular}{cc}
        \includegraphics{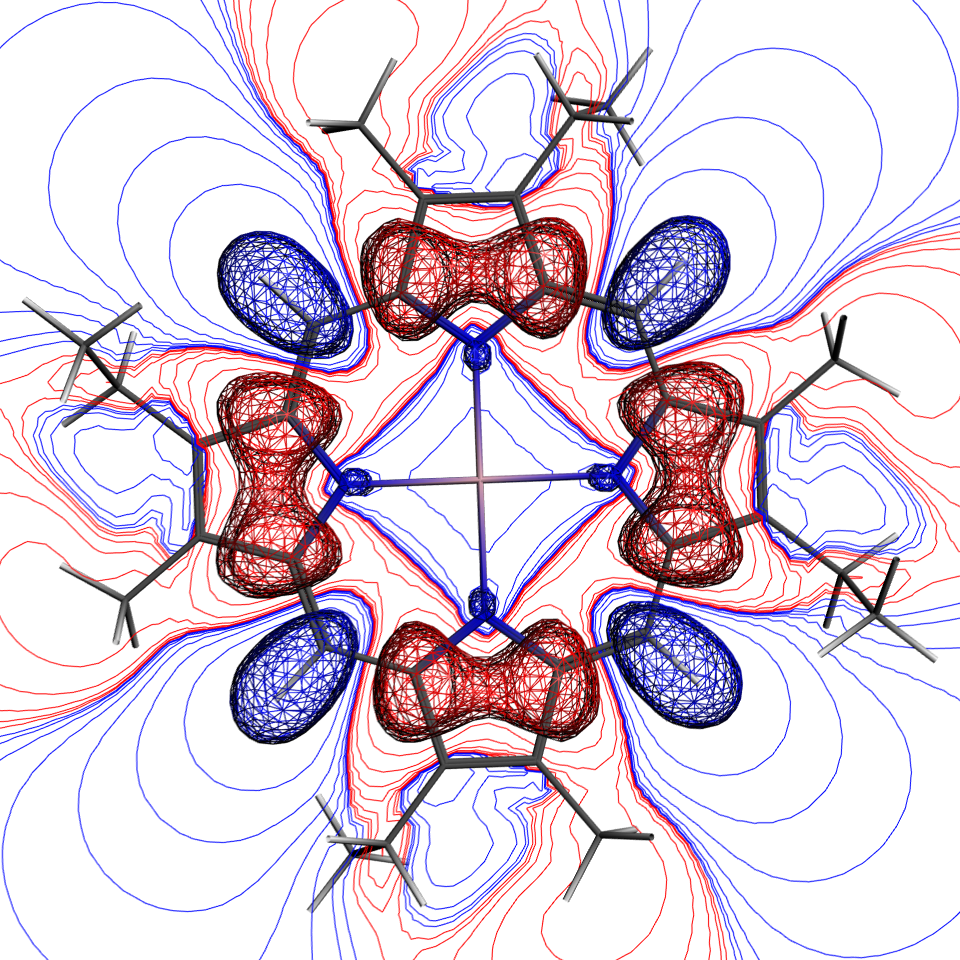} &
        \includegraphics{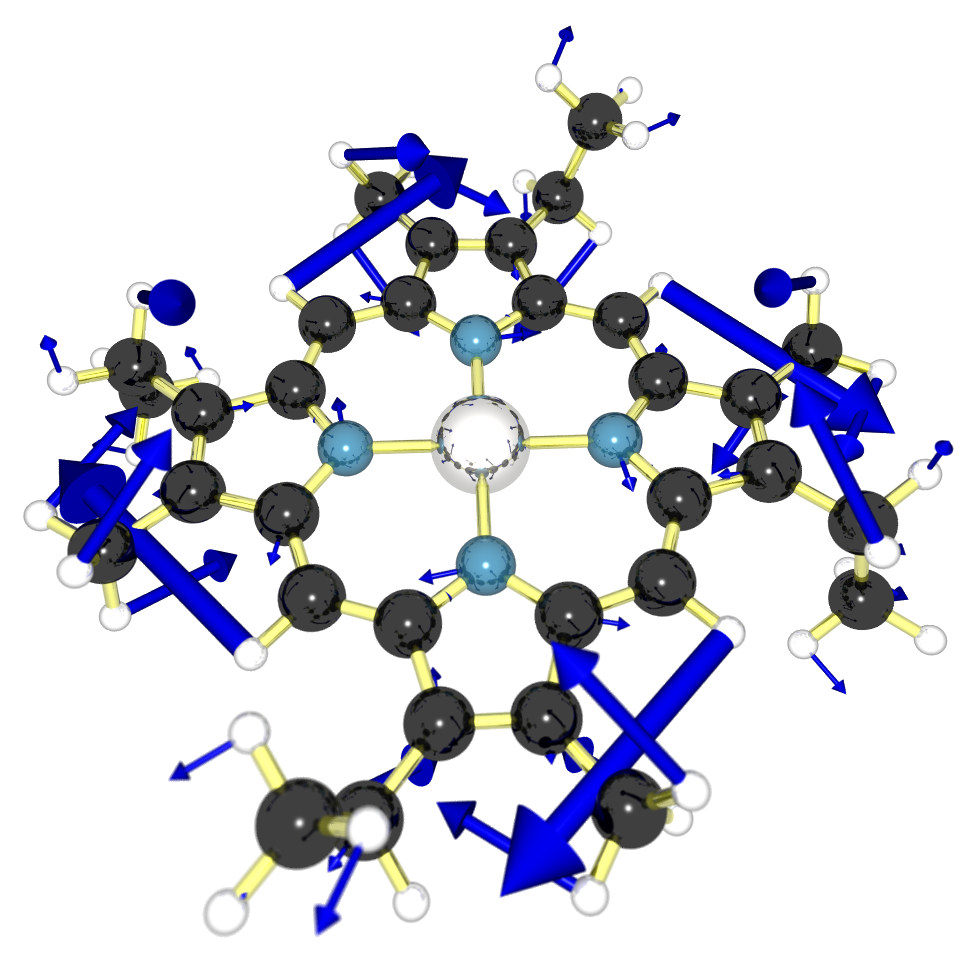} \\
        (a) & (b) 1319 cm$^{-1}$ \\
        \includegraphics{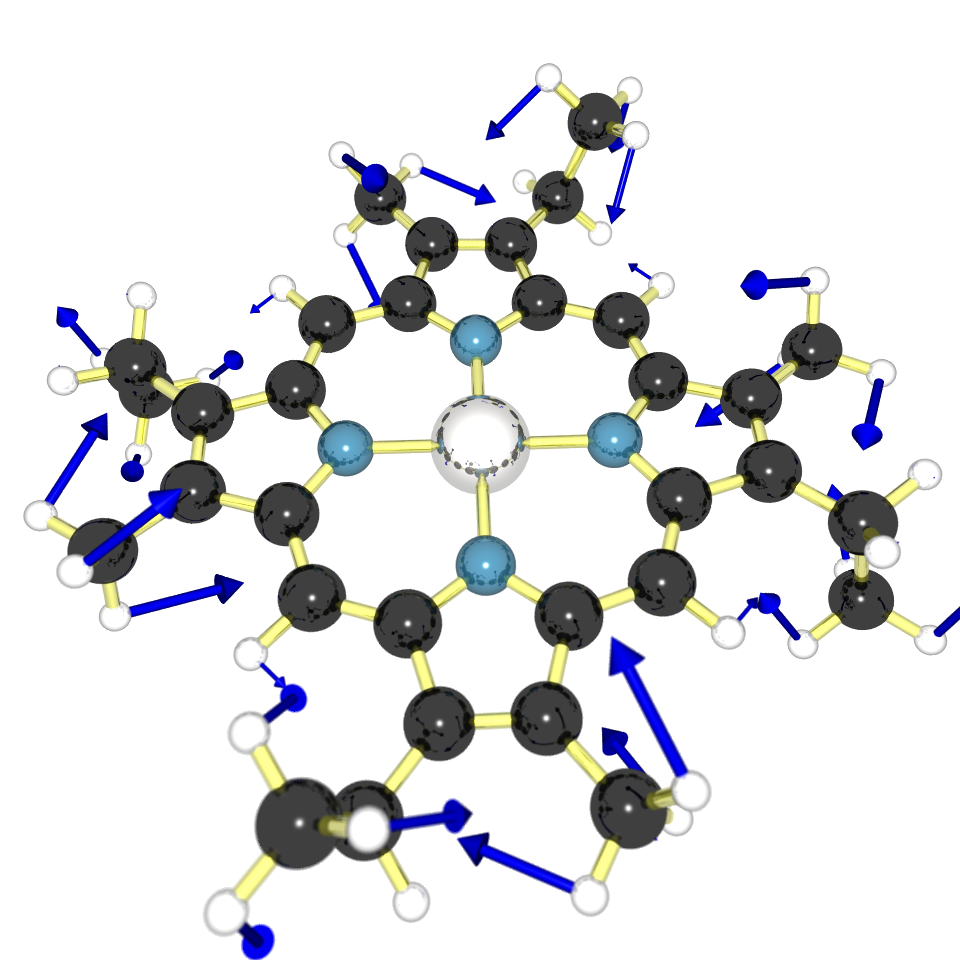} &
        \includegraphics{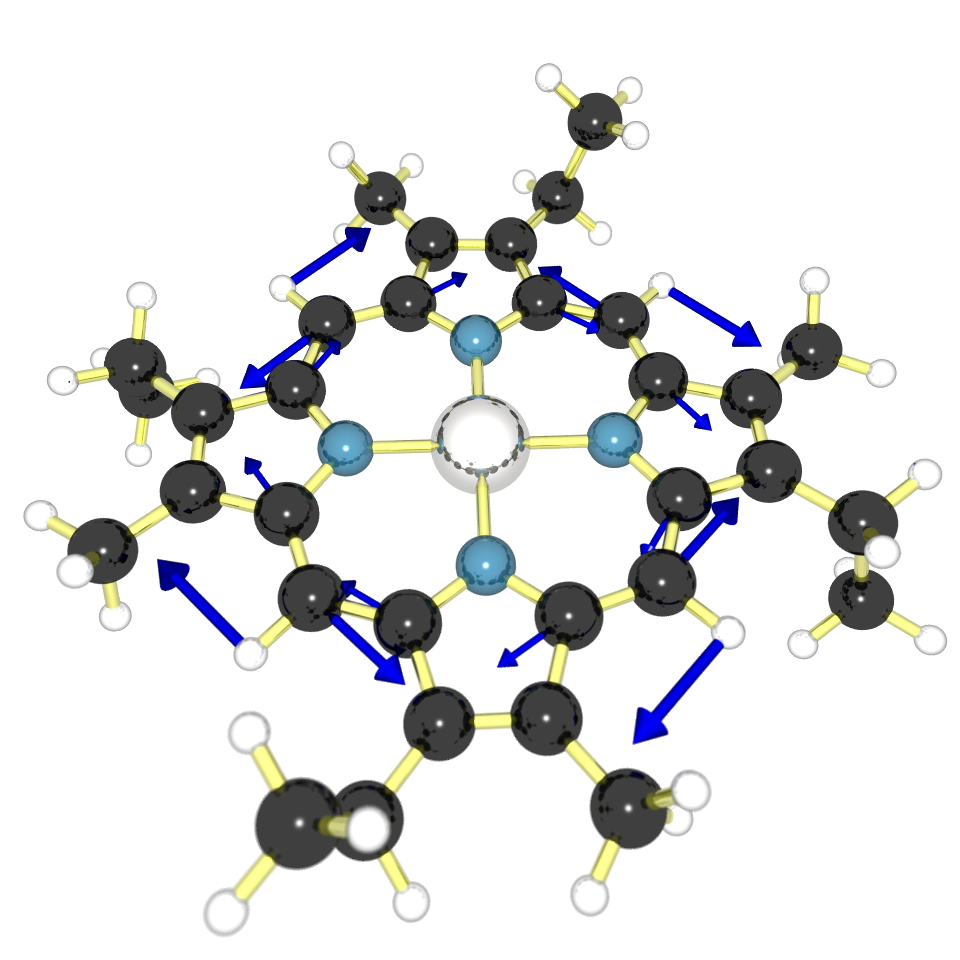} \\
        (c) 1337 cm$^{-1}$ & (d) 1614 cm$^{-1}$ \\
    \end{tabular}
    \caption{\label{fig:ZnEtioI_modes} (a) Calculated change in the Coulomb potential felt by the ZnEtioI nuclei due to the emission of a photon. [(b)--(d)] The three most important vibrational modes responsible for the peaks around 900~nm in Fig.~\ref{fig:ZnEtioI_spectrum}(b). The motions of the nuclei can be directly related to the potential gradients in (a).}
\end{figure}

The measured and calculated bias voltage and current dependence of the electroluminescence spectrum is shown in Fig.~\ref{fig:ZnEtioI_spectrum}. The gradual increase in the photo-emission intensity at higher bias voltages in the measurement is not clearly reproduced in our rate-based calculation, since level broadening is not taken into account. Comparison of the calculated spectrum to the measurement, on the other hand, shows good agreement. Both spectra show several small side peaks beyond the main peak at 790~nm and a series of peaks around 900~nm. The temperature of the experimental setup (77~K) makes the identification of individual vibrational modes difficult, but the calculation indicates that only a few of the 213 modes are active in the electroluminescence (see Fig.~\ref{fig:ZnEtioI_modes}). The peak at 790~nm is dominated by the vibrational ground-state to ground-state transition, while the peaks at 900~nm consist of a handful of modes involving pyrrole breathing and twisting modes and rotations of the methyl/ethyl side groups. The most important modes can be seen in Figs.~\ref{fig:ZnEtioI_modes}(b)--\ref{fig:ZnEtioI_modes}(d). The reason for these modes to be active in the electroluminescence spectrum becomes clear when looking at the change in the Coulomb potential due to the emission of a photon [Fig.~\ref{fig:ZnEtioI_modes}(a)]. This change is defined as the change in the electrostatic potential felt by the nuclei due the difference in the electron charge density between the ground state and the excited state. The gradient of the potential, and therefore the force, is largest where the potential suddenly changes sign [between the red and blue areas in Fig.~\ref{fig:ZnEtioI_modes}(a)]. Comparing the potential gradient with the most important vibrational modes shows that those atoms move which are close to a large potential gradient.

Compared to the measurement, the peaks at 900~nm are lower with respect to the main peak at 790~nm than they are in the measurement. This may in part be caused by the electroluminescence background due to the NiAl substrate around that wavelength, or a varying sensitivity of the charge-coupled-device camera in the spectral range, but is most likely mainly caused by the limited number of vibrational quanta taken into account in the calculation,\cite{Note4} resulting in the transitions at higher wavelengths to be under-represented in the spectrum.

In a different measurement, Qiu \emph{et al.}\ report observing equidistant vibrational features with a peak spacing of 40$\pm$2~meV [Fig.~5C in Ref.~\onlinecite{Qiu2003}]. They suggest these peaks are possibly higher harmonics of the same vibrational mode. In the calculation, however, none of the vibrational modes has a sufficiently large electron-phonon coupling\cite{Note5} to produce such a ladder. The calculation does show a series of \emph{different} vibrational modes with a non-zero electron-phonon coupling spaced approximately 40~meV apart.

Besides a dependence on the voltage, Qiu \emph{et al.}\cite{Qiu2003} also find a linear dependence of the photon count on the current [see the inset in Fig.~\ref{fig:ZnEtioI_spectrum}(a)]. This linear dependence is reproduced in our calculations [Fig.~\ref{fig:ZnEtioI_spectrum}(c)] and can be easily understood. In the case of asymmetric couplings, where the source electrode (STM tip) is much more weakly coupled than the drain (substrate), the average occupation of the excited state is nearly zero: it will take a long time for an electron to tunnel onto the molecule, but once it is there, it will tunnel off almost immediately. The coupling of the source electrode ($\Gamma_\text{S}$) can be varied by changing the vertical position of the STM tip. This changes both the current through the molecule and the photo-emission rate, as the latter is directly proportional to the average occupation of the excited state. Since the ratio between the photo-emission rate and the electron transmission rate is independent of $\Gamma_\text{S}$ ($\approx\tfrac{1}{5}\frac{\Gamma_\text{E}}{\Gamma_\text{D}}$, see Sec.~\ref{sec:Method}), this results in a linear dependence of the photon count on the current. Note that linearly changing the current by varying $\Gamma_\text{S}$ is only possible when $\Gamma_\text{S}\ll\Gamma_\text{D}$.

\subsection{H$_2$TBPP}\label{sec:H2TBPP}

\begin{figure}
    \begin{tabular}{cc}
        \includegraphics{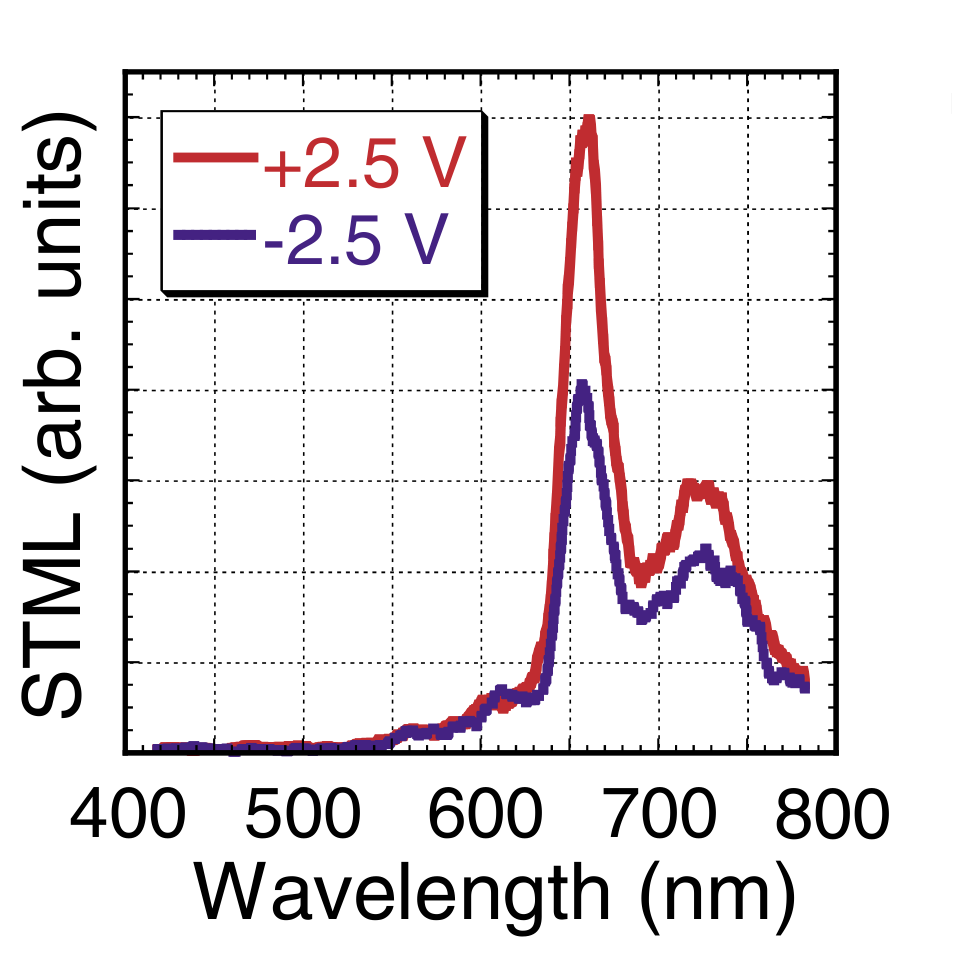} &
        \includegraphics{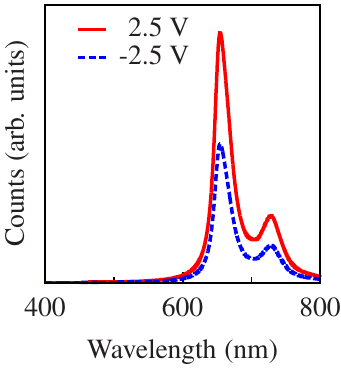} \\
        (a) & (b)
    \end{tabular}
    \caption{\label{fig:H2TBPP_spectrum} (a) Measured electroluminescence spectrum of H$_2$TBPP at positive and negative bias. [Reprinted figure with permission from Dong \emph{et al.}\ (Ref.~\onlinecite{Dong2004}) copyright (2004) by the American Physical Society.] (b) Calculated electroluminescence spectrum (assuming instantaneous vibrational relaxation). The electronic coupling is nearly symmetric ($\Gamma_\text{S}=16.4$~$\mu$eV and $\Gamma_\text{D}=4.8$~$\mu$eV) and $T=300$~K.}
\end{figure}

\begin{figure}
    \begin{tabular}{cc}
        \includegraphics{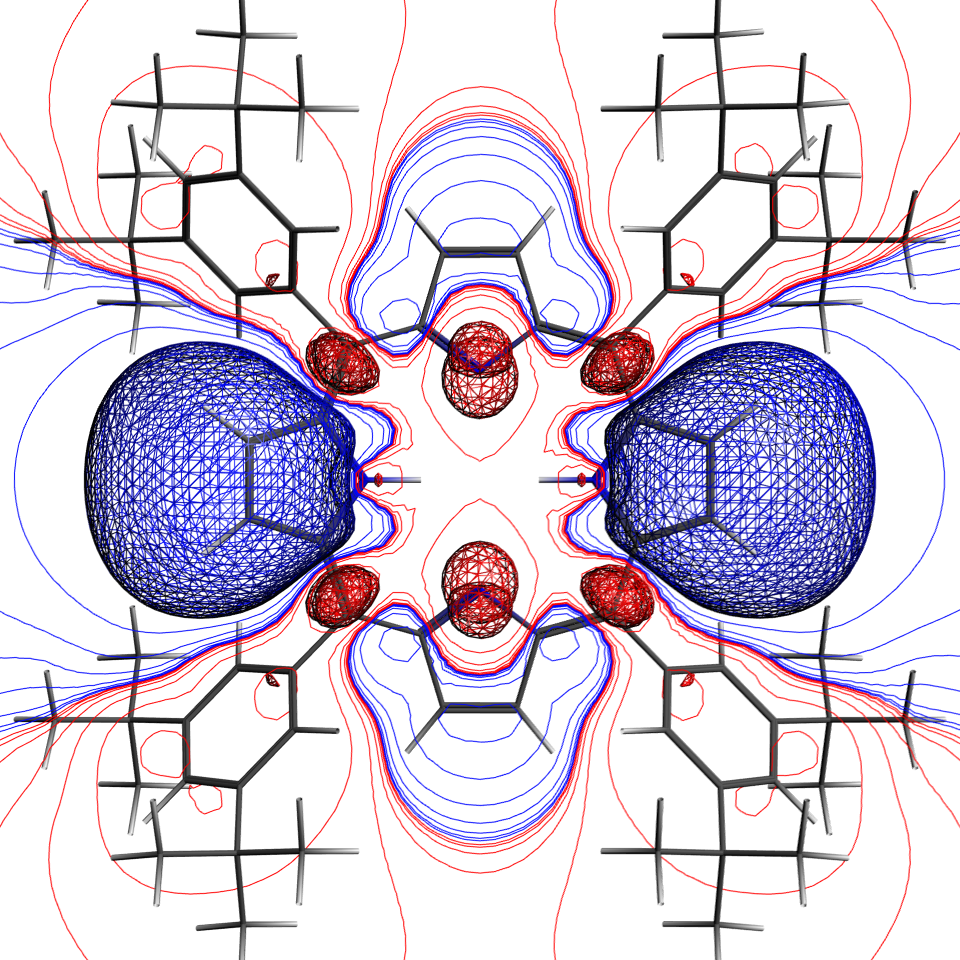} &
        \includegraphics{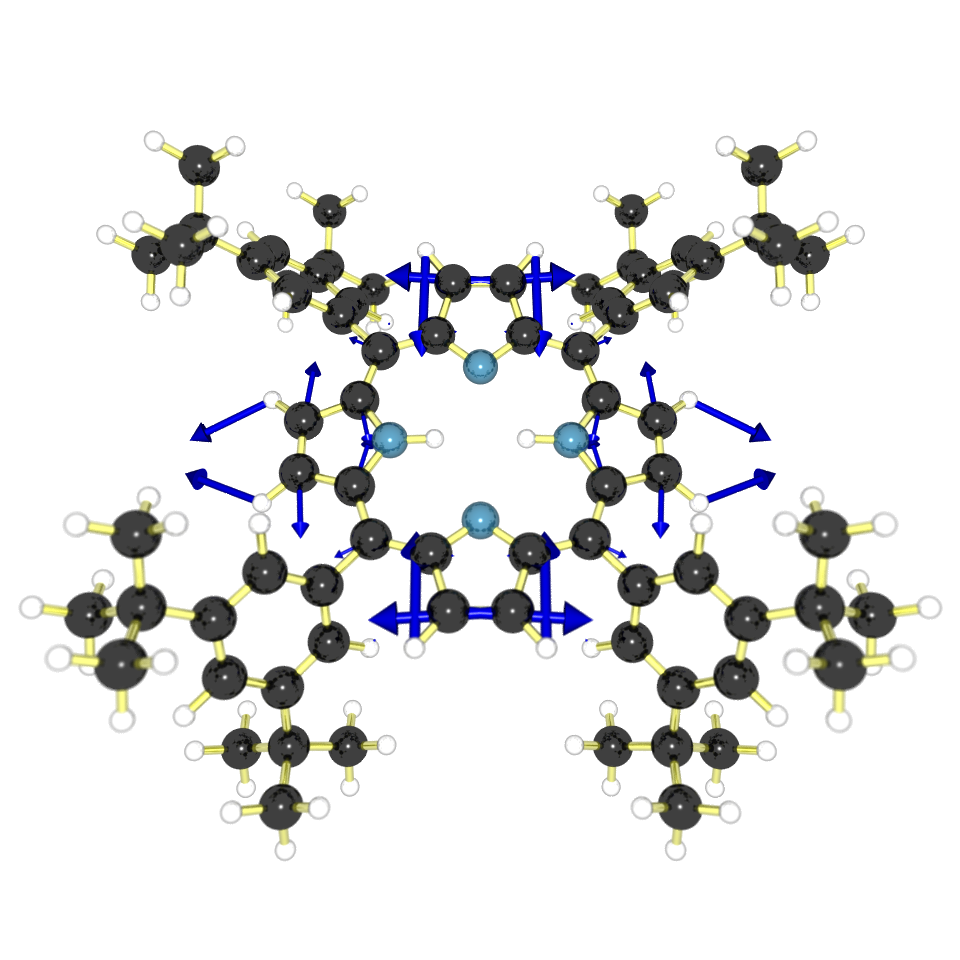} \\
        (a) & (b) 1530 cm$^{-1}$ \\
        \includegraphics{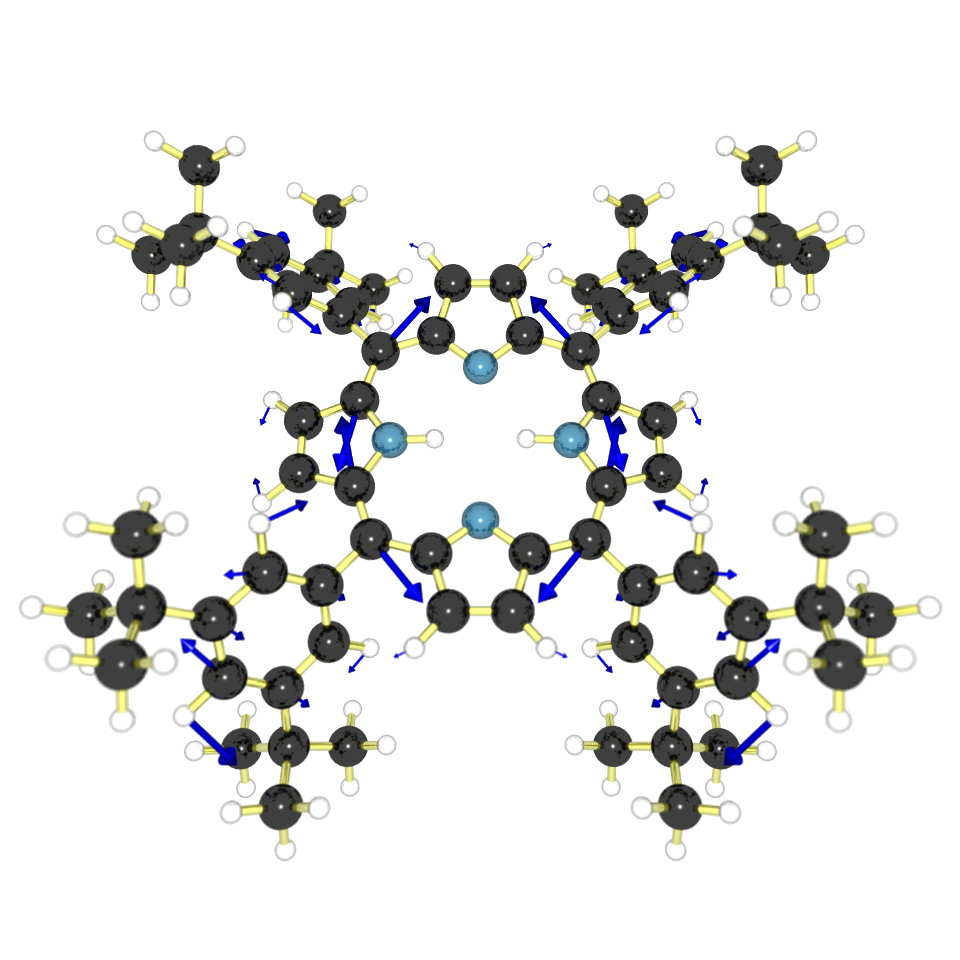} &
        \includegraphics{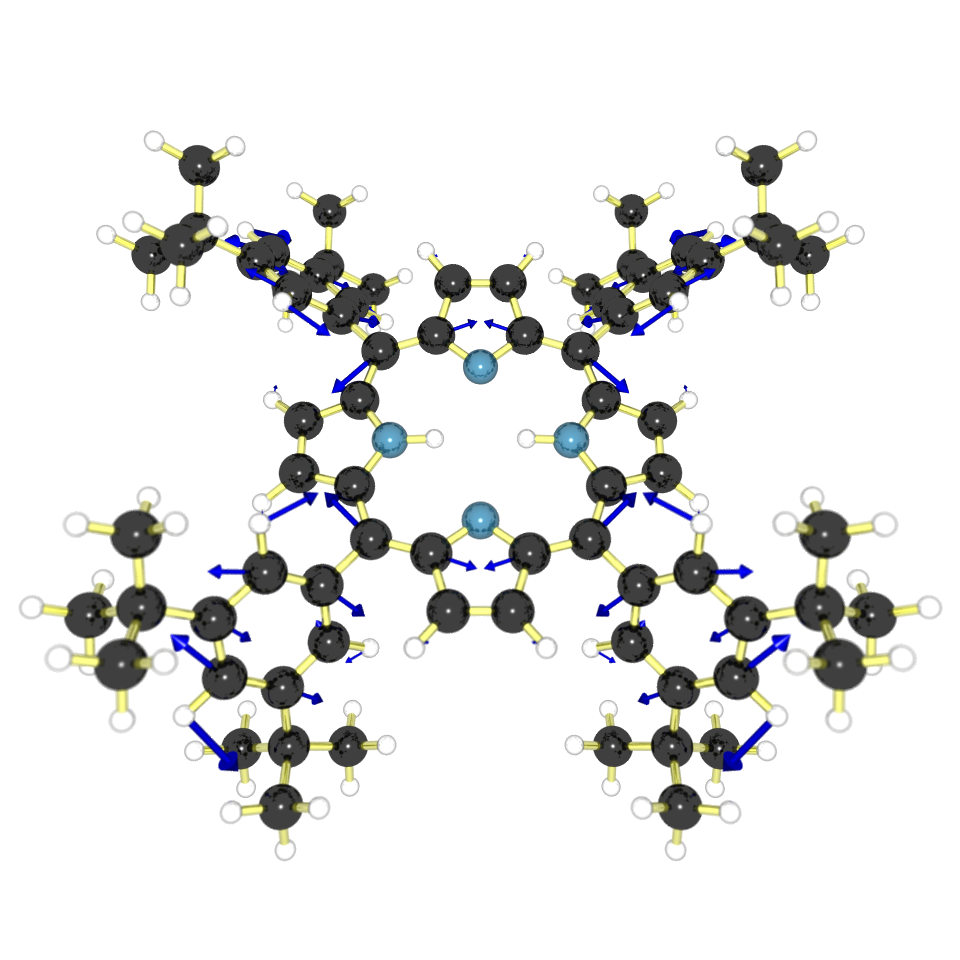} \\
        (c) 1612 cm$^{-1}$ & (d) 1625 cm$^{-1}$ \\
    \end{tabular}
    \caption{\label{fig:H2TBPP_modes} (a) Calculated change in the Coulomb potential felt by the H$_2$TBPP nuclei due to electroluminescence from the LUMO. The potential change for the LUMO+1 is similar, but rotated by 90$^\circ$. [(b)--(d)] The three most important vibrational modes responsible for the peaks around 723~nm in Fig.~\ref{fig:H2TBPP_spectrum}(b).}
\end{figure}

In the measurements of Dong \emph{et al.},\cite{Dong2004} the molecule is coupled almost symmetrically to the leads. The equilibrium configuration of the H$_2$TBPP molecule used in the calculation is shown in Fig.~\ref{fig:structure}(b). The $\mathcal{C}_{\text{2v}}$ conformation, and, in particular, the angle between the porphyrin center and the four side groups, is in agreement with the STM measurements of Jung \emph{et al.}\cite{Jung1997}

The measured and calculated electroluminescence spectra at positive and negative bias are shown in Fig.~\ref{fig:H2TBPP_spectrum}. Note that electroluminescence at both positive and negative bias is only expected when the molecule is (nearly) symmetrically coupled to the leads (see below). The small asymmetry in the coupling causes the photon count at negative bias to be approximately half of that at positive bias. As in the measurement of Qiu \emph{et al.},\cite{Qiu2003} the spectrum consists of a main peak around 658~nm, corresponding to the vibrational ground state to ground-state transition, and then another peak at around 723~nm. As with the ZnEtioI calculation, the lower-energy transitions are somewhat under-represented in the calculated spectrum. The temperature of the experimental setup (300~K) again makes the identification of individual modes impossible, but the calculation predicts only a few active modes around the energy of the peak at 723~nm. These correspond to pyrrole breathing and twisting modes --- explaining why spectra of the two molecules are so similar --- and to rotations of the side groups (see Fig.~\ref{fig:H2TBPP_modes}).

\begin{figure}
    \begin{tabular}{cc}
        \includegraphics{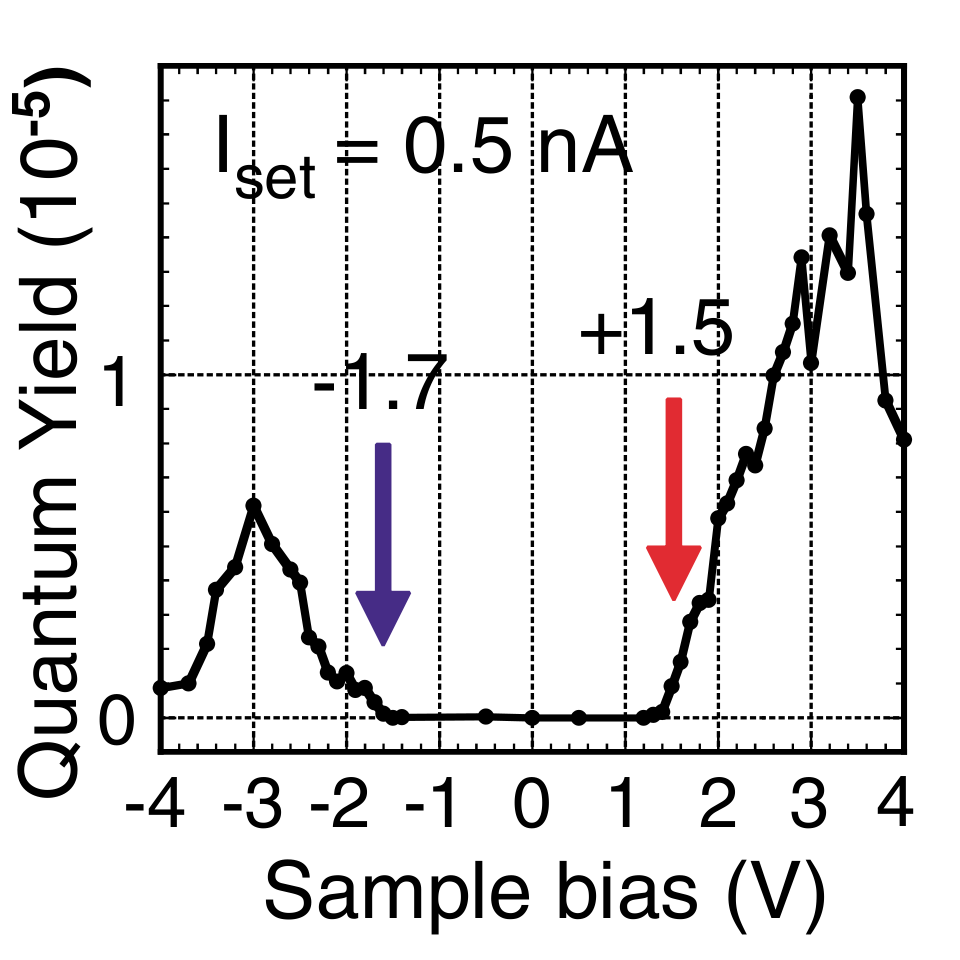} &
        \includegraphics{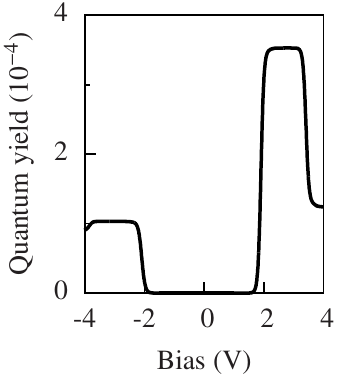} \\
        (a) & (b)
    \end{tabular}
    \caption{\label{fig:H2TBPP_qc} (a) Measured quantum yield of H$_2$TBPP as a function of bias voltage. [Reprinted figure with permission from Dong \emph{et al.}\ (Ref.~\onlinecite{Dong2004}) copyright (2004) by the American Physical Society.] (b) Calculated quantum yield, assuming a 100~\% detection efficiency of the emitted photons.}
\end{figure}

The measured and calculated quantum yield as a function of the bias voltage are shown in Fig.~\ref{fig:H2TBPP_qc}. The difference in absolute magnitude between the measurement and the calculation is due to the unknown detection efficiency in the measurement, which is taken to be a 100~\% in the calculation. In both the measurement and the calculation, the quantum yield is zero until an excited state becomes available in the bias voltage window.\cite{Note6} It then increases until a maximum is reached when all vibrationally excited states have become available. As noted above, the quantum yield depends on $\Gamma_\text{D}$, and not on $\Gamma_\text{S}$. Since going from positive to negative bias effectively means switching $\Gamma_\text{D}$ and $\Gamma_\text{S}$, the maximum value of the quantum yield changes by a factor of $\frac{\Gamma_\text{S}}{\Gamma_\text{D}}$, and is therefore directly proportional to the asymmetry in the electronic coupling.

In the measurement, the quantum yield drops when the bias exceeds 3.5~V, which is attributed to damage to the molecules in Ref.~\onlinecite{Dong2004}. This can, however, also be explained by the appearance of another charge state in the bias window, which changes the quantum yield by a factor of $\frac{\Gamma_\text{D}}{1.2\Gamma_\text{S}+\Gamma_\text{S}}$ (see Sec.~\ref{sec:Method}). With approximately equal couplings to the source and drain ($\Gamma_\text{S}\approx\Gamma_\text{D}$), the quantum yield is reduced by about 50\% [as is the case in Fig.~\ref{fig:H2TBPP_qc}(b)]. However, with very asymmetric, couplings ($\Gamma_\text{S}\ll\Gamma_\text{D}$), as is the case in the measurement of Qiu \emph{et al.},\cite{Qiu2003} the reduction is expected to be unobservable.

\begin{figure}
    \includegraphics{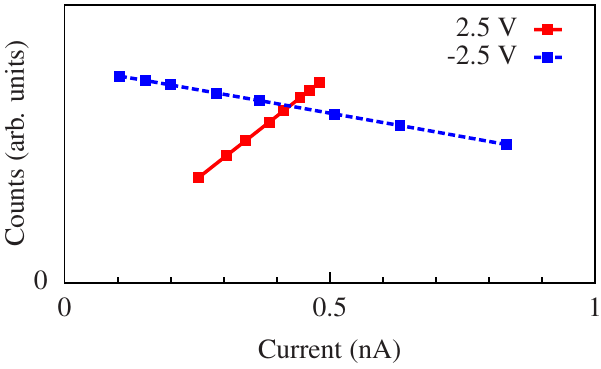}
    \caption{\label{fig:H2TBPP_current} Calculated current dependence of H$_2$TBPP at positive and negative bias. The current is varied by changing the electronic coupling to the STM tip.}
\end{figure}

Dong \emph{et al.}\cite{Dong2004} do not report on the current dependence of the photon count.\cite{Note7} However, in the case of symmetric coupling, a linear dependence of the photon count on the current, by varying $\Gamma_\text{S}$ with the STM tip, is still expected at positive bias (see Fig.~\ref{fig:H2TBPP_current}). At negative bias, on the other hand, varying the position of the STM tip equals changing $\Gamma_\text{D}$, resulting in a \emph{decrease} in the photon count with increasing current, albeit at a lower rate. This is a direct result of the dependence of the quantum yield on $\Gamma_\text{D}$ via $Q\approx\tfrac{1}{5}\frac{\Gamma_\text{E}}{\Gamma_\text{D}}$ (see Sec.~\ref{sec:Method}).

\begin{figure}
    \includegraphics{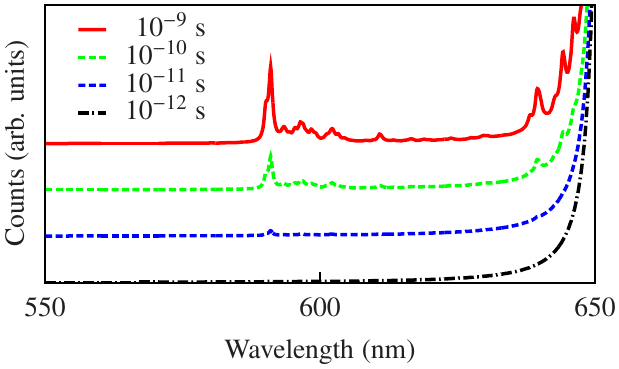}
    \caption{\label{fig:H2TBPP_relaxation} High-energy part of the calculated electroluminescence spectrum of H$_2$TBPP at 2.5~V as a function of the vibrational relaxation time. $T=15$~K for clarity, the other parameters are the same as in Fig.~\ref{fig:H2TBPP_spectrum}(b). Note the appearance of peaks at around 600~nm, to the left of the main peak, when the vibrational relaxation time becomes comparable to the electron transmission rate.}
\end{figure}

Since vibrational relaxation is several orders of magnitude faster than photo-emission (which is on the order of 10$^{-6}$~s in the measurements and calculations under discussion), it is assumed in standard fluorescence measurements that the molecule is always in the vibrational ground state before emitting a photon, although at finite temperatures the lowest vibrational excitations can still be occupied, \emph{e.g.}, as observed in anti-Stokes scattering. In transport measurements, however, even the higher vibrationally excited states are continuously repopulated by new electrons from the leads. When the electron transmission rate becomes on the order of the photo-emission rate, vibrational relaxation will not always take place before the electron leaves the excited state. It is therefore the ratio between the vibrational relaxation rate and the electron transmission rate, not just the photo-emission rate, that determines the emission spectrum. This can be seen in Fig.~\ref{fig:H2TBPP_relaxation}, where we have plotted the high energy part of the emission spectrum of H$_2$TBPP for different values of the vibrational relaxation time. Peaks start appearing to the \emph{left} of the main peak, \emph{i.e.}, at higher energies, as soon as the vibrational relaxation rate becomes comparable to the electron transmission rate (at around 10$^{-10}$~s). This effect may be present in the electroluminescence spectrum measured by Dong \emph{et al.}\cite{Dong2004} [see Fig.~\ref{fig:H2TBPP_spectrum}(a)], where such a peak can be seen at around 40~nm to the left of the main peak.

\section{Conclusions}

In conclusion, we have shown that a computationally efficient method --- all calculations have been performed on commodity hardware --- based on the rate-equation formalism is able to obtain good agreement with the measured electroluminescence spectra of two different porphyrin derivatives. We reproduce the bias voltage and current dependence of the spectra for both symmetric and asymmetric couplings and provide an explanation for the suppression of the quantum yield at high bias. Although in the measurement of Dong \emph{et al.}\cite{Dong2004} this suppression may have been caused by damage to molecules,\cite{Dong2004} a significant reduction in the quantum yield is expected in general for every (nearly) symmetrically coupled single-molecule junction in the sequential tunneling regime. In addition, we have shown that vibrational relaxation rates become important when they are comparable to the electron transmission rate, giving rise to peaks in the spectra at higher energies higher than the HOMO-LUMO gap. However, a detailed study of this effect requires a more sophisticated model for vibrational relaxation, since a single relaxation time for all vibrational modes is only a sufficiently accurate approximation when relaxation is either much faster or much slower than all other rates.

\begin{acknowledgments}
We thank Ferry Prins for discussions and Chris Verzijl for help with the calculations. Financial support was obtained from Stichting FOM (project 86), from the EU FP7 program under the grant agreement ``SINGLE'', and from the Division of Chemistry and the Division of Materials Research of the NSF through the Northwestern University MRSEC.
\end{acknowledgments}

\input{manuscript.bbl}

\end{document}